\documentclass[amsmath,amssymb,pdftex,aps,10pt,rmp,twocolumn,superscriptaddress,letterpaper]{revtex4}
\usepackage{graphicx}
\usepackage{color}

\usepackage[english]{babel}
\usepackage[T1]{fontenc}
\usepackage{textcomp}

\usepackage[activate={true,nocompatibility},final,tracking=true,kerning=true,spacing=true,factor=1100,stretch=10,shrink=10]{microtype}
\microtypecontext{spacing=nonfrench}

\usepackage{makecell}
\usepackage{mathptmx}
\usepackage{float}
\usepackage{setspace}
\usepackage{tabularx}
\usepackage[sc,tiny,noindentafter]{titlesec}
\titleformat{\subsection}[hang]{\itshape}{}{}{}

\definecolor{darkgray}{rgb}{0.25,0.25,0.25}
\definecolor{darkred}{rgb}{0.89,0.10,0.11}
\definecolor{darkblue}{rgb}{0.12,0.39,0.62}
\usepackage{url}

\usepackage[pdftex,breaklinks=true,colorlinks=true,citecolor=black,linkcolor=black,menucolor=black,urlcolor=darkblue,pdfborder={1 0 0}]{hyperref}
\hypersetup{pdftitle={Infomap Bioregions: Interactive mapping of biogeographical regions from species distributions},pdfauthor={D. Edler, T. Guedes, A. Zizka, M. Rosvall and A. Antonelli (2016)}}

\usepackage{booktabs}

\begin{document}
	
\renewcommand{\figurename}{Figure}
\renewcommand{\thefigure}{\arabic{figure}}
\renewcommand{\tablename}{Table}
\renewcommand{\thetable}{\arabic{table}}
\renewcommand{\refname}{\large References}
\renewcommand\refname{references}

\setlength{\belowcaptionskip}{1ex}
\setlength{\textfloatsep}{2ex}
\setlength{\dbltextfloatsep}{2ex}

\title{\texorpdfstring{\normalfont\sc\normalsize INFOMAP BIOREGIONS:\\ interactive mapping of biogeographical regions from species distributions}{Infomap Bioregions: Interactive mapping of biogeographical regions from species distributions}}


\author{\normalfont Daniel Edler}
\email{daniel.edler@umu.se}
\affiliation{\normalfont  Integrated Science Lab, Department of Physics, Ume{\aa} University, SE-901 87 Ume{\aa}, Sweden}
\affiliation{\normalfont  University of Gothenburg, Department of Biological and Environmental Sciences, Box 461, SE-405 30 G\"{o}teborg, Sweden}

\author{\normalfont Tha\'{i}s Guedes}
\affiliation{\normalfont University of Gothenburg, Department of Biological and Environmental Sciences, Box 461, SE-405 30 G\"{o}teborg, Sweden}
\affiliation{\normalfont Federal University of S\~{a}o Paulo, 09972-270, Diadema, Brazil}
\affiliation{\normalfont Museum of Zoology of University of S\~{a}o Paulo, 04263-000, S\~{a}o Paulo, Brazil}

\author{\normalfont Alexander Zizka}
\affiliation{\normalfont University of Gothenburg, Department of Biological and Environmental Sciences, Box 461, SE-405 30 G\"{o}teborg, Sweden}

\author{\normalfont Martin Rosvall}
\affiliation{\normalfont Integrated Science Lab, Department of Physics, Ume{\aa} University, SE-901 87 Ume{\aa}, Sweden}

\author{\normalfont Alexandre Antonelli}
\affiliation{\normalfont University of Gothenburg, Department of Biological and Environmental Sciences, Box 461, SE-405 30 G\"{o}teborg, Sweden}
\affiliation{\normalfont Gothenburg Botanical Garden, Carl Skottsbergs gata 22A, SE-413 19 Gothenburg, Sweden}

\begin{abstract}
\normalsize\rule{0cm}{4ex}
Biogeographical regions (bioregions) reveal how different sets of species are spatially grouped and therefore are important units for conservation, historical biogeography, ecology and evolution. Several methods have been developed to identify bioregions based on species distribution data rather than expert opinion. One approach successfully applies network theory to simplify and highlight the underlying structure in species distributions. However, this method lacks tools for simple and efficient analysis. Here we present Infomap Bioregions, an interactive web application that inputs species distribution data and generates bioregion maps.
Species distributions may be provided as georeferenced point occurrences or range maps, and can be of local, regional or global scale. The application uses a novel adaptive resolution method to make best use of often incomplete species distribution data. The results can be downloaded as vector graphics, shapefiles or in table format. We validate the tool by processing large datasets of publicly available species distribution data of the world's amphibians using species ranges, and mammals using point occurrences. We then calculate the fit between the inferred bioregions and WWF ecoregions. As examples of applications, researchers can reconstruct ancestral ranges in historical biogeography or identify indicator species for targeted conservation.\\
\end{abstract}

\maketitle

\section*{introduction}
Biodiversity is not randomly distributed. It is well known that  species are grouped in space and patterns of distribution can be recognized at small and large scales. Depending on the size, source of data and scientific discipline, these broadly used biogeographical regions have received various related names, but here we simply refer to them as bioregions (see \citet{vilhena2015network} for a discussion of terminology). In many disciplines, working with bioregions rather than single species is more effective. Conservation biology is a prime example, since protecting bioregions with high levels of biodiversity or uniqueness may help protecting many species from extinction. In historical biogeography, bioregions may be used as operational areas for ancestral range reconstructions in order to estimate how lineages in a phylogeny have evolved their geographical occupancy over time \citep{goldberg2011phylogenetic,matzke2014model,ree2008maximum}.
Moreover, since different taxa exhibit different patterns of diversity, distribution, and evolutionary history, there is no set of universal bioregions for all circumstances. Accordingly, the most effective set of bioregions depends on the particular system under study and research question at hand. Therefore, researchers need simple, effective, and flexible tools for mapping relevant species distribution data into bioregions.

While bioinformatic tools can now provide rapid and accurate coding of species into predefined areas \citep{topel02082016}, choosing the areas in the first place has been a subjective procedure without quantitative support.  Researchers have therefore developed a suite of algorithms for mapping grid cell areas into biologically relevant regions \citep{kozak2006does,kreft2010framework,oliveira2015delimiting}, but often they involve multiple and overly technical steps. As a consequence, most biogeographical studies still use arbitrarily defined areas.

To make identification of bioregions simple and effective for any set of species distribution data, we present the web-based, interactive mapping tool Infomap Bioregions. The underlying method clusters bipartite networks that contain both species and grid cells. This method was recently shown to outperform approaches that abstract away the species into species similarities between grid cells in unipartite networks \citep{vilhena2015network}. Moreover, the bipartite networks are clustered with the information-theoretic clustering algorithm known as Infomap \citep{rosvall2008maps}, which has been acclaimed as the best network clustering algorithm in several comparative studies \citep{Lancichinetti2009,Aldecoa2013}. Thanks to its simple and effective design, Infomap Bioregions has wide applications in biogeography, ecology, evolutionary biology, conservation and related disciplines.

\section*{Description}
\begin{figure}[]
  \centering
  \includegraphics[width=\columnwidth]{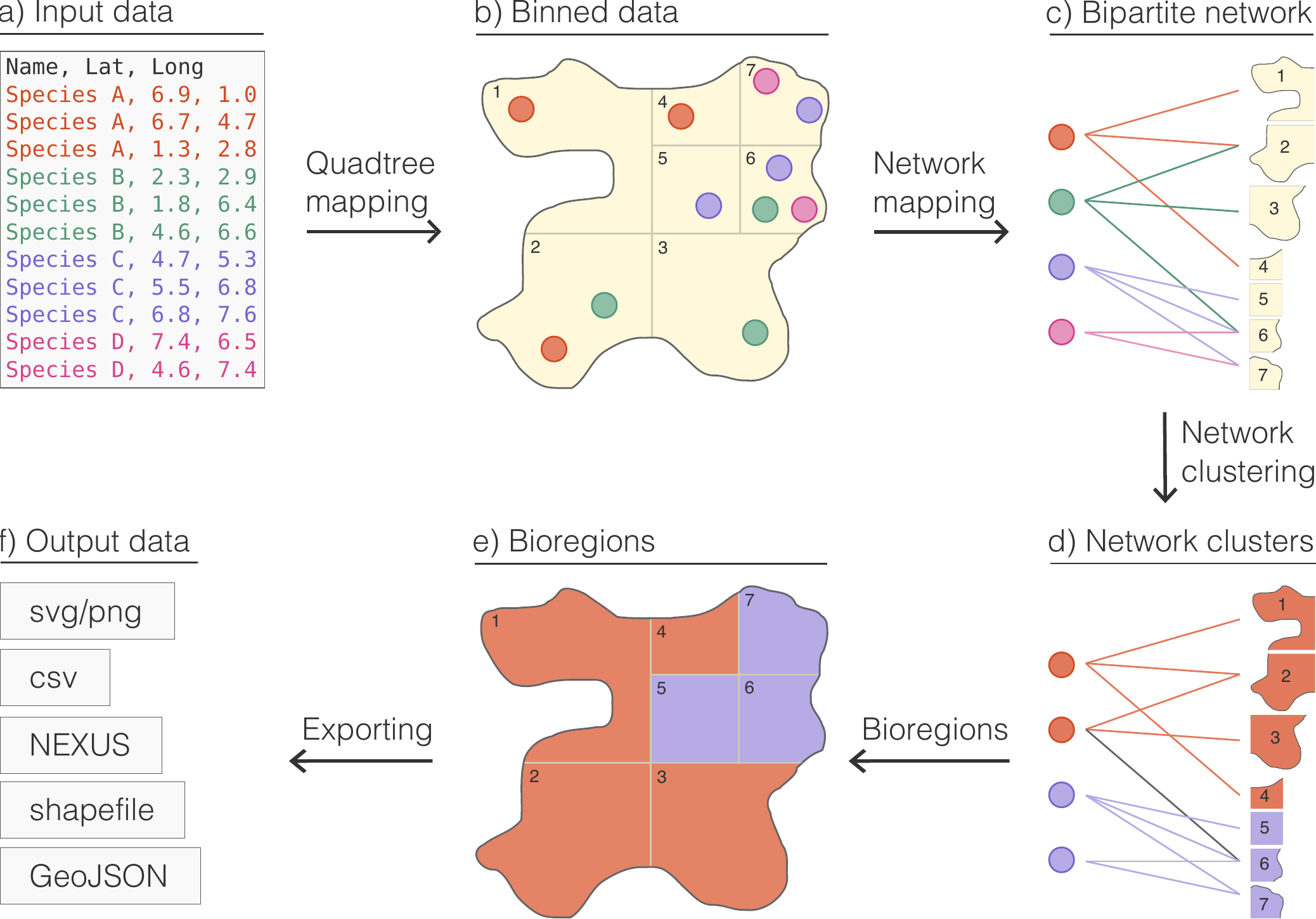}
  \caption{Step-by-step illustration of how Infomap Bioregions generates bioregions from species distribution data. Infomap Bioregions: a) inputs comma-separated values for point occurrences, b) adaptively bins species records into discrete geographical grid cells such that the data density determines the spatial resolution, c) extracts a  bipartite network between species and grid cells, d) clusters the bipartite network with the \emph{Infomap clustering algorithm}, e) visualises the grid cell clusters as bioregions on a zoomable map, f) exports the geographical map in \texttt{svg} or \texttt{png} format, the tables of top occurring and top indicative species for each bioregion in \texttt{csv} format, the species presence/absence matrix for the bioregions in \texttt{NEXUS} format and the geographical information of the bioregions in \texttt{shapefile} or \texttt{GeoJSON} format.}
  \label{fig:schematics}
\end{figure}

Infomap Bioregions is an interactive web application that identifies taxon-specific bioregions from species distribution data. We first present the application's workflow (Fig.\ \ref{fig:schematics}), and then describe each step in detail.

Given user-provided species distribution data, the application first bins the data into geographical grid cells with adaptive spatial resolution. When the data are sparse, the grid size is large; and when the data are dense, the grid size is small. This novel adaptive resolution offers a considerable advantage over conventional uniform binning when dealing with biodiversity data, which is unevenly distributed \citep{maldonado2015estimating, meyer2016multidimensional}.

The binning generates a bipartite network between species and grid cells, which is then clustered with the Infomap algorithm into bioregions \citep{infomap}. The application also identifies the most common and the most indicative species in each grid cell and bioregion. The results are shown as an interactive map together with supporting tables containing information about each bioregion.

To facilitate the integration of bioregion delimitation and ancestral range reconstructions, Infomap Bioregions also supports loading a phylogenetic tree, which may be time-calibrated or not. Fitch's method of maximum parsimony as originally described \citep{fitch1971toward} is implemented to provide a quick estimate of ancestral ranges. Species in the phylogeny that are not present in the distribution dataset are ignored in the ancestral range reconstruction, and ancestral ranges for the remaining species are shown with pie charts based on the bioregions identified (see Fig.\ \ref{fig:mammals}). The bioregions can also be exported to allow analyses under alternative methods of ancestral range reconstruction.

\subsection*{Input data}
\label{sub:Input data}

For species distribution data, Infomap Bioregions supports both point occurrences and species range maps. Point occurrences are specified in a text file with either comma-separated (CSV) or tab-separated (TSV) values. The application requires a header with the column names, and the user must identify which columns that should be parsed as name, latitude and longitude, respectively (Fig.\ \ref{fig:schematics}a). Range maps are specified in the shapefile format, which includes multiple files: a \texttt{.shp} file for species range polygons, a \texttt{.dbf} file for the attributes of each range polygon and, optionally, a \texttt{.prj} file for projection information. As for point occurrence data, the user must identify which attribute to parse as the name of the species.

For phylogenetic data, Infomap Bioregions supports the NEXUS and Newick tree formats.

\subsection*{Output data}
\label{sub:Output data}
The map with bioregions can be exported in \texttt{.svg} and \texttt{.png} format. The shapes of bioregions can be exported in \texttt{.geojson} and \texttt{shapefile} format, and a species presence/absence matrix for the bioregions can be exported in \texttt{.nexus} format for further analyses in e.g. BioGeoBEARS \citep{matzke2013biogeobears}, BayArea \citep{landis2013bayesian}, and many other ancestral reconstruction tools that can handle such files.

Summary tables of the most common and the most indicative species for each bioregion can be exported in \texttt{.csv} format, and the tree can be exported in \texttt{.svg} format.

\subsection*{Adaptive resolution}
\label{sub:Adaptive resolution}
Where data are sparse, single cells can be clustered in distinct bioregions. To avoid providing more detail than the data can support, Infomap Bioregions automatically adapts the grid size to the amount and spatial distribution of the input data. This is done by mapping the input data to a so-called quadtree data structure, as illustrated in Figure \ref{fig:schematics}b.

The adaptive resolution algorithm uses the quadtree to hierarchically partition geographical space into quadratic grid cells of increasingly smaller size by recursively subdividing each grid cell into four quadrants. When the algorithm reaches the user-provided \emph{maximum cell size} (default is $4^{\circ}$), it aggregates the species into grid cells.

To make the resolution adaptive to the density of the data, each grid cell has a user-provided \emph{maximum cell capacity} (default is 100 species occurrence records). The algorithm recursively subdivides all grid cells with more records than the maximum cell capacity until it reaches the user-provided \emph{minimum cell size} (default is $1^{\circ}$). However, if a grid cell after a subdivision contains less species than the user-provided \emph{minimum cell capacity} (default is 10), the algorithm reverts the most recent subdivision to avoid creating regions with too few data points.

With these criteria, Infomap Bioregions can simultaneously identify high-resolution bioregions where data are abundant and low-resolution bioregions where data are sparse, and thereby avoid over- and underfitting across all bioregions.

For point occurrence data, these criteria make the adaptive resolution  straightforward.
For range maps, the application first adds a species record to each grid cell of minimum size that intersects with the corresponding species range polygon, and then proceeds with the adaptive binning to satisfy the user-specified criteria.

It is also possible to interactively modify the resolution of the bioregions by adjusting the Markov time for the Infomap clustering algorithm \citep{kheirkhah2016efficient}. In this way, the user can tune Infomap to search for bigger or smaller bioregions that are still supported by the data.

\subsection*{Bipartite network}
\label{sub:Bipartite network}
When Infomap Bioregions aggregates the species into geographical grid cells, it forms a bipartite network with \emph{species} and \emph{grid cells} as the two types of nodes. Each species is connected by an unweighted link to each grid cell in which it is present. We purposefully avoid weighting the links by the number of records, because that would make the results sensitive to spatially biased sampling. Instead, we let the density of species records increase the spatial resolution as described above. In this way, dense data generate large networks.

\subsection*{Bioregions and indicator species}
\label{sub:Bioregions and indicator species}

Infomap Bioregions clusters the bipartite network with Infomap for bipartite networks \citep{kheirkhah2016efficient}. The resulting clusters contain both grid cells and species, and define the bioregions. The software displays the bioregions with different colours on a map, and provides a table for each bioregion including summary statistics and species lists.
The application lists for grid cells and bioregions both the most common species and the most indicative species with the highest relative abundance. That is, for species $s$ in grid cell or bioregion $r$, the indicative score $I_{s|r}$ is defined as the ratio between the frequency $f_{s|r}$ of the species in the region and the frequency $f_s$ of the species in all regions, $I_{s|r} = f_{s|r}/f_s$. Thus an indicative score of 2 means that a species is twice as frequent in the region than in the entire dataset. In the bioregions tables, the most common and indicative species are displayed together with charts that show the distribution of those species in other bioregions. This information makes it possible to find endemic species, unique or close to unique to a specific bioregion.

\section*{Results and discussion}

To validate Infomap Bioregions, we applied it to range maps of amphibians and point occurrences of terrestrial mammals. For amphibians, we downloaded global distribution data as range polygons for 6,069 species from the International Union for Conservation of Nature (IUCN, \url{http://www.iucn.org}, downloaded August 17, 2015, from \url{http://www.iucnredlist.org/technical-documents/spatial-data}). For terrestrial mammals, we compiled the global distribution of 5,005 species from a collection of georeferenced observation records obtained through the Global Biodiversity Information Facility (\url{GBIF.org} (11th November 2015; GBIF Occurrence Download \url{http://doi.org/10.15468/dl.wnjjkc}). We cleaned the mammal dataset using the R functions in the package \texttt{speciesgeocodeR} \citep{topel02082016}, checking for obvious errors such as empty coordinates,terrestrial species reported in the sea, and coordinates assigned to country or province centroids.

For the resolution, we allowed grid cells to range between $4^{\circ}$ and $2^{\circ}$ to reflect spatial differences in data density. We used maximum cell capacity 100 and set the minimum cell capacity to 5. Below we show the bioregion maps of the amphibians and mammals, and highlight a few bioregions.

For terrestrial mammals, we downloaded a set of 1000 species-level phylogenies of all mammals from \citep{faurby2015species} and calculated the maximum clade credibility tree using TreeAnnotator in the package BEAST v.1.8.2 \citep{drummond2007beast}. We were able to match 4,426 species between the phylogeny and the georeferenced dataset.

\subsection*{Amphibians}
\label{subs:Amphibians}

We identified 87 bioregions of amphibians as illustrated in Figure \ref{fig:amphibians} (see Supplementary Material available on Dryad \url{http://dx.doi.org/10.5061/dryad.2s201} for detailed results).
In Table\ \ref{table:amphibians}, we detail the three most species rich bioregions and three small bioregions with relatively many species. Most of the species belong to relatively large bioregions, but we also identified smaller bioregions such as in the Caribbean where island endemics are common and in the tropical Andes where species turnover is high and many species are located in just a few cells.
\begin{figure*}[htp]
  \centering
  \includegraphics[width=\textwidth]{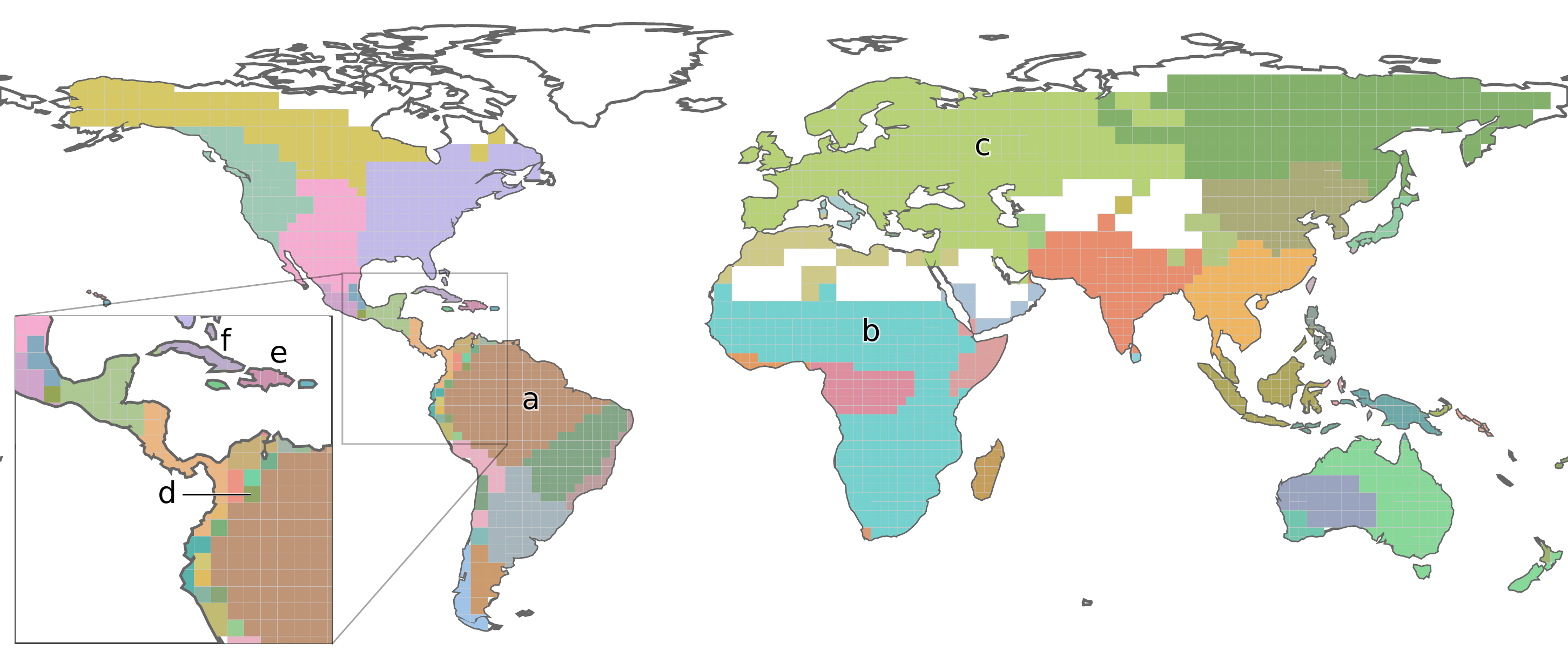}
  \caption{Bioregion map of the world's amphibians generated with Infomap Bioregions, using the IUCN species range maps. White areas have insufficient data and were excluded from the analysis. The inset shows a zoom-in of Central America, the West Indies and northwestern South America, depicting many small bioregions that reflect high turnover of species assemblages and narrow range distributions characteristic for the region. Table \ref{table:amphibians} shows information about labelled bioregions.}
  \label{fig:amphibians}
\end{figure*}
\renewcommand{\arraystretch}{0.8}
\setlength{\tabcolsep}{1em}
\begin{table*}[htp]
\centering
\caption{Selected amphibian bioregions. For exact locations, the indices (a)--(f) are displayed on the bioregion map in Figure \ref{fig:amphibians}. Bioregions (a)--(c) are the most species rich and (d)--(f) are hand-picked to illustrate how even small bioregions can contain relatively many species. Common names from Encyclopedia of Life at \url{http://eol.org}}
\label{table:amphibians}
\begin{tabular}{@{}llllll@{}}
\toprule
Location      & Records                    & Species                 & Cells                   & Most common species  (records)                    & Most indicative species (score)            \\ \midrule
(a) South America & \multicolumn{1}{r}{42,161} & \multicolumn{1}{r}{719} & \multicolumn{1}{r}{167} & \textit{Trachycephalus venulosus} (600) & \textit{Lithobates palmipes} (3.3)  \\
&  &  &  & Veined tree frog & Amazon River frog \\ \noalign{\smallskip}
(b) Africa & \multicolumn{1}{r}{27,267} & \multicolumn{1}{r}{553} & \multicolumn{1}{r}{333} & \textit{Kassina senegalensis} (970)     & \textit{Hildebrandtia ornata} (2.1) \\
&  &  &  & Senegal running frog & African ornate frog \\ \noalign{\smallskip}
(c) Eurasia & \multicolumn{1}{r}{13,083} & \multicolumn{1}{r}{103} & \multicolumn{1}{r}{313} & \textit{Rana arvalis} (1,547)     & \textit{Triturus cristatus} (1.3) \\
&  &  &  & Moor frog & Northern crested newt \\ \noalign{\smallskip}
(d) Andes & \multicolumn{1}{r}{121} & \multicolumn{1}{r}{75} & \multicolumn{1}{r}{1} & \textit{Pristimantis nervicus} (13)     & \textit{Pristimantis nervicus} (157) \\
&  &  &  & -- & -- \\ \noalign{\smallskip}
(e) Hispaniola & \multicolumn{1}{r}{181} & \multicolumn{1}{r}{65} & \multicolumn{1}{r}{4} & \textit{Hypsiboas heilprini} (28)     & \textit{Osteopilus vastus} (73) \\
&  &  &  & Los Bracitos tree frog & Hispaniola tree frog \\ \noalign{\smallskip}
(f) Cuba & \multicolumn{1}{r}{214} & \multicolumn{1}{r}{61} & \multicolumn{1}{r}{4} & \textit{Osteopilus septentrionalis} (28)     & \textit{Eleutherodactylus varleyi} (73) \\
&  &  &  & Cuban tree frog & -- \\ \bottomrule
\end{tabular}
\end{table*}

The identified bioregions largely coincide with those found by \citet{vilhena2015network}, except for some differences due to the adaptive resolution and its settings. For the Neotropics, our clustering seems to reflect the regionalization proposed by  \citet{morrone2006biogeographic,morrone2014biogeographical} for some sub-regions and provinces such as the Amazonian subregion, the Parana subregion and the Chacoan dominion. Infomap Bioregions also successfully identified small bioregions, for example in the island of Hispaniola and in the tropical Andes, which could be particularly considered for conservation. Other examples of relatively small-scale bioregions include the Cape region in South Africa and the Dahomey gap in West Africa (Fig.\ \ref{fig:amphibians}).

\subsection*{Mammals}
\label{subs:Mammals}

We identified 62 bioregions of mammals, which we show together with their phylogenetic tree and ancestral range reconstructions in Figure \ref{fig:mammals} (see Supplementary Material available on Dryad \url{http://dx.doi.org/10.5061/dryad.2s201} for detailed results). Some of the bioregions are very large, reflecting major continental-wide differences, while others comprise no more than a few square degrees. For example, we identified more than 10 bioregions for Australia, a landmass known to contain a high number of species and ecosystems (see Table \ref{table:mammals}). 

\begin{figure*}[btp]
  \centering
  \includegraphics[width=\textwidth]{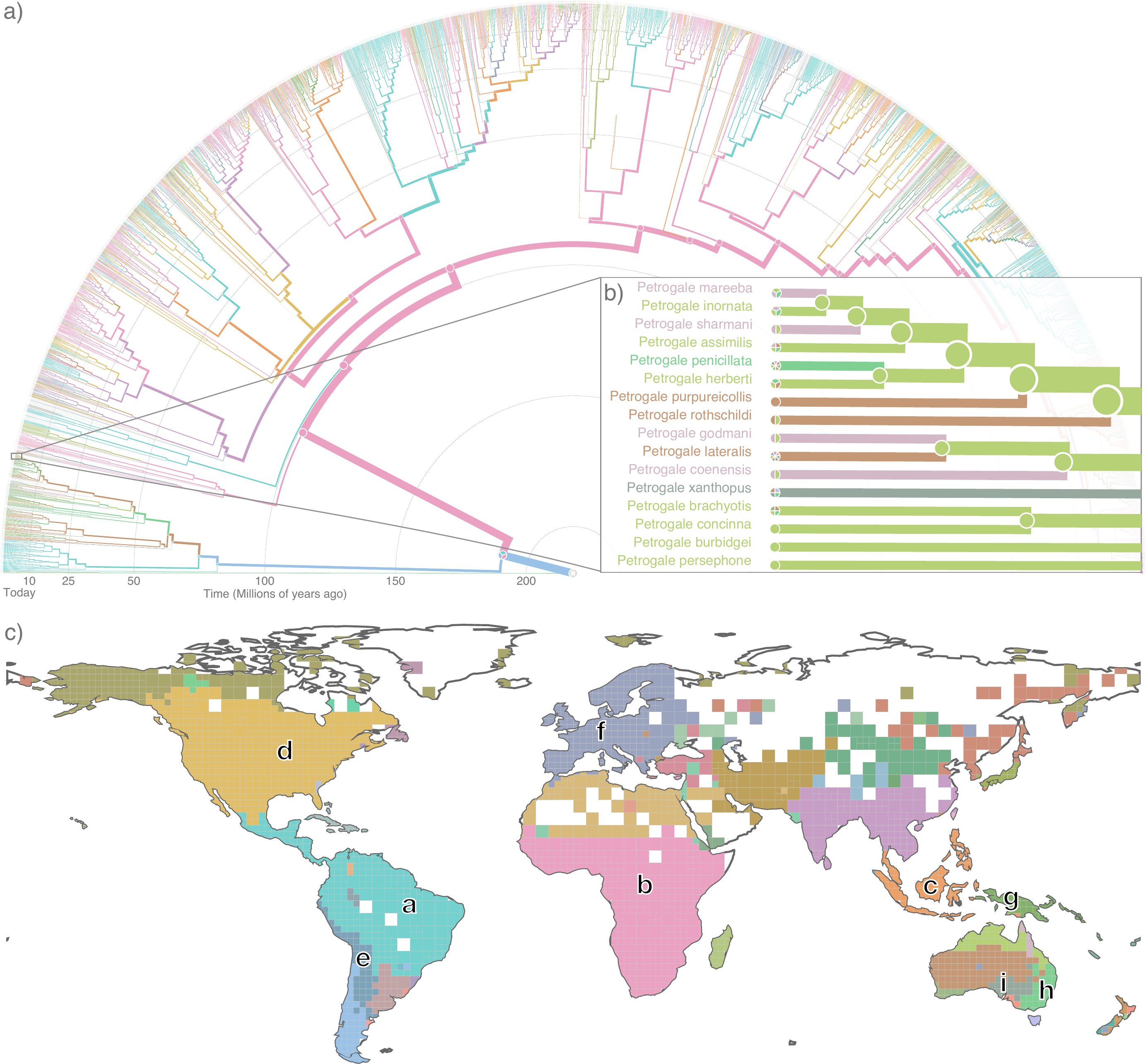}
  \caption{Bioregion map and phylogenetic tree of world mammals with ancestral range reconstruction, generated with Infomap Bioregions. a) Phylogenetic tree of 5,747 mammals computed from \citep{faurby2015species}, fully zoomable on the online application. Ancestral ranges reconstructed under Fitch parsimony. Pie charts depict most parsimonious ancestral ranges at nodes, and current distributions for extant species. Branch lines are scaled to the number of terminals subtending each branch, in order to improve visualisation of the overall tree structure. b) Magnified part of the tree, highlighting the rock-wallabies (genus \emph{Petrogale}) which are currently distributed across several bioregions in Australia. This analysis suggests that all rock-wallabies, including the Yellow-footed rock-wallaby (\emph{Petrogale xanthopus}), which is the most indicative species of the southeast bioregion (i), originated from a common ancestor in northern Australia. c) Bioregion map of world mammals using species point occurrences from GBIF. White areas have insufficient data and were excluded from the analysis. Colours are used consistently across the subfigures. Table \ref{table:mammals} shows information about labelled bioregions.}
  \label{fig:mammals}
\end{figure*}

\setlength{\tabcolsep}{0.75em}
\begin{table*}[htp]
\centering
\caption{Highlighted mammalian bioregions, sorted on species richness. For exact locations, the indices (a)--(i) are displayed on the bioregion map in Figure \ref{fig:mammals}. Common names from Encyclopedia of Life at \url{http://eol.org}}
\label{table:mammals}
\begin{tabular}{@{}llllll@{}}
\toprule
Location      & Records                    & Species                 & Cells                   & Most common species (records)                     & Most indicative species (score)            \\ \midrule
(a) South America & \multicolumn{1}{r}{69,757} & \multicolumn{1}{r}{1,448} & \multicolumn{1}{r}{254} & \textit{Glossophaga soricina} (1,588)     & \textit{Uroderma bilobatum} (36) \\
&  &  &  & Pallas's long-tongued bat & Tent-making Bat \\ \noalign{\smallskip}
(b) Africa & \multicolumn{1}{r}{38,258} & \multicolumn{1}{r}{1,105} & \multicolumn{1}{r}{323} & \textit{Mastomys natalensis} (991)     & \textit{Mus musculoides} (58) \\
&  &  &  & Common African rat & Kasai mouse \\ \noalign{\smallskip}
(c) Malay Archipelago & \multicolumn{1}{r}{8,788} & \multicolumn{1}{r}{576} & \multicolumn{1}{r}{103} & \textit{Rattus exulans} (370)     & \textit{Ptenochirus jagori} (155) \\
&  &  &  & Polynesian rat & Greater Musky Fruit Bat \\ \noalign{\smallskip}
(d) North America & \multicolumn{1}{r}{279,416} & \multicolumn{1}{r}{812} & \multicolumn{1}{r}{426} & \textit{Peromyscus maniculatus} (17,600)     & \textit{Thomomys bottae} (3.3) \\
&  &  &  & Deer mouse & Valley Pocket Gopher \\ \noalign{\smallskip}
(e) Andes & \multicolumn{1}{r}{4,709} & \multicolumn{1}{r}{394} & \multicolumn{1}{r}{46} & \textit{Phyllotis xanthopygus} (280)     & \textit{Akodon albiventer} (189) \\
&  &  &  & Yellow-rumped leaf-eared mouse & White-bellied grass mouse \\ \noalign{\smallskip}
(f) Europe & \multicolumn{1}{r}{635,148} & \multicolumn{1}{r}{389} & \multicolumn{1}{r}{266} & \textit{Meles meles} (46,299) & \textit{Talpa europaea} (1.2)  \\
&  &  &  & Eurasian badger & European mole \\ \noalign{\smallskip}
(g) New Guinea & \multicolumn{1}{r}{5,107} & \multicolumn{1}{r}{325} & \multicolumn{1}{r}{46} & \textit{Syconycteris australis} (260)     & \textit{Echymipera kalubu} (220) \\
&  &  &  & Southern blossom bat & Common Echymipera \\ \noalign{\smallskip}
(h) SE Australia & \multicolumn{1}{r}{298,374 } & \multicolumn{1}{r}{269} & \multicolumn{1}{r}{39} & \textit{Phascolarctos cinereus} (26,029)     & \textit{Petaurus australis} (2.2) \\
&  &  &  & Koala & Yellow-bellied glider \\ \noalign{\smallskip}
(i) SE Australia & \multicolumn{1}{r}{43,669} & \multicolumn{1}{r}{147} & \multicolumn{1}{r}{22} & \textit{Macropus robustus} (9,176)     & \textit{Petrogale xanthopus} (6.2) \\
&  &  &  & Hill wallaroo & Yellow-footed rock-wallaby \\
\bottomrule
\end{tabular}
\end{table*}

We acknowledge that the automated cleaning steps described above for species occurrences are probably not sufficient to fully validate the distribution dataset. Careful revision of specimens and localities by taxonomists, and increased spatial sampling, are some of the time-consuming tasks required to produce more reliable datasets \citep{maldonado2015estimating, meyer2016multidimensional}. As a consequence, our results may be affected by sampling biases, inaccurate georeferencing, and/or incorrect identifications. These issues prevent us from discerning, for example, whether the scattered occurrence of small bioregions in e.g.\ Russia is a real biological result or, more likely, an artefact of the scarce publicly available data for that region.

\subsection*{Validation}
\label{subs:Validation}
\begin{figure*}[btp]
  \centering
  \includegraphics[width=\textwidth]{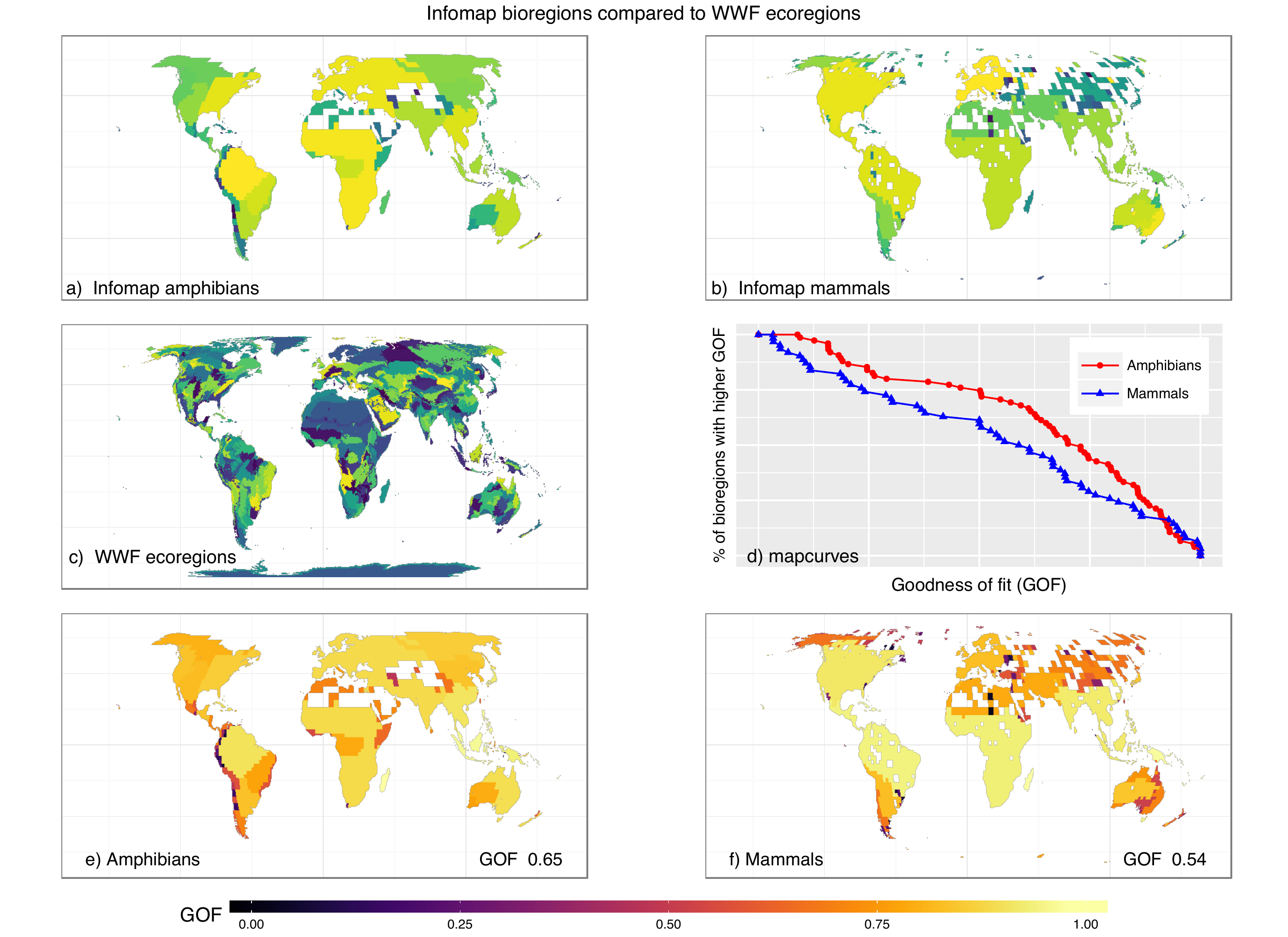}
  \caption{Comparison between Infomap bioregions and WWF ecoregions using the Mapcurves algorithm \citep{hargrove2006mapcurves}. a) Infomap bioregions for the amphibian dataset; b) Infomap bioregions for the mammalian dataset; c) WWF ecoregions from \citep{olson2001terrestrial}; d) Mapcurves as a measure for the goodness of fit (GOF) for the Infomap bioregions with respect to the WWF ecoregions. The graph shows the percentage of bioregions with a GOF score better than the corresponding value on the horizontal axis (zero to one). A perfect fit for all bioregions would be indicated by a horizontal line at the top; e) GOF map of the Infomap bioregions for amphibians and the total GOF score; f) GOF map of the Infomap bioregions for mammals and the total GOF score. The fit of the bioregions to the WWF ecoregions is generally very good, with the exception of several very small bioregions identified by Infomap Bioregions.}
  \label{fig:validation}
\end{figure*}

We further evaluated the performance of Infomap Bioregions by comparing the bioregions identified for mammals and amphibians with the widely used WWF ecoregions \citep{olson2001terrestrial}, see Figure \ref{fig:validation}. For these analyses we used the mapcurves algorithm \citep{hargrove2006mapcurves} as implemented by \citep{mapcurves2011implementation}. Mapcurves is a quantitative method to compare the spatial concordance between categorical maps, by calculating a goodness of fit (GOF) for each polygon in a map of interest based on the degree of spatial overlap with the polygons of a reference map. The results can be summarized in a global GOF score. Mapcurves is resolution independent, does not require the same number of categories in both maps, and any polygon in a map that can be exactly comprised of a set of polygons in the reference map will show a perfect fit. However, the algorithm generally indicates a poor fit when a finer resolution map is compared to a coarser map. A limitation of this asymmetry becomes apparent when the map of interest has coarser resolution than the chosen reference map in some or most of the areas, but finer resolution in other areas. Therefore, the globally best GOF map may have areas of finer resolution where the local GOF is poor, whereas the bioregions in the reference map covering the same area may have better local GOF. See \citet{hargrove2006mapcurves} for a detailed description of the method.

The results of the comparison show a generally good fit between the bioregions identified by Infomap Bioregions and the WWF ecoregions. The fit was better for amphibians (global GOF 0.65) than for mammals (0.54), which might be related to the the data used for map creation (range polygons vs. point occurrences, respectively). The spatial visualization of the GOF scores shows that the fit is very good for most areas. Differences are mainly associated with a number of very small bioregions, mostly in the Andes for amphibians and mostly in Asia and northern Africa for mammals. For mammals, many of the small bioregions recognized by Infomap Bioregions seem to derive from low data availability. A low fit is also partly an artefact of the asymmetry of the measure as mentioned above, especially for amphibians in the Andes.

We also compared the bioregions identified by Infomap Bioregions to the zoogeographic regions from \citep{holt2013update}, which for the amphibians were based on approximately the same data, but delimited using a beta-diversity method including phylogenetic information. The GOF was overall very good, with a global GOF score of 0.77 for amphibians and 0.52 for mammals (see  Supplementary Material). In summary, the results of the pairwise comparisons show that bioregions delimited by Infomap Bioregions generally correspond well to commonly used bioregionalization maps, despite  differences in the underlying data and methodology applied. 

\section*{Conclusions}

Designed to make data-driven identification of bioregions simple and effective, we introduced the web application Infomap Bioregions and demonstrated its flexibility. A user can load species data from both point occurrences and range polygons, modify parameters directly in the web interface, and export results to various formats for high-quality printing or further biogeographical analyses. The web application uses adaptive spatial resolution, can process millions of records in a few minutes, and applies bipartite network clustering that outperforms traditional methods based on similarity indices. Moreover, the user can load phylogenetic data for the species and explore how the bioregions map to the phylogenetic tree. We validated the application on two large datasets of amphibians and mammals and anticipate that Infomap Bioregions will become a standard tool in many studies in ecology, evolution, conservation biology and historical biogeography.

\section*{Availability and forthcoming extensions}
\label{Availability}

Infomap Bioregions is made open source under the GNU AGPL v3\texttt{+} license. It is written in JavaScript and builds on a set of open source libraries (see dependencies in \texttt{package.json}). Because it is a pure client-side application, all data stay and all calculations run on the user's computer. Moreover, all heavy calculations run in a background thread. Together, this means improved privacy and performance.

Infomap Bioregions is available at \url{http://bioregions.mapequation.org} and the source code is freely available at \url{http://github.com/mapequation/bioregions}.

Possible forthcoming extensions include batch runs, additional methods to find indicator species and bioregions, hierarchical clustering of bioregions, deeper integration of phylogenetic information and significance clustering with bootstrap to find which bioregional boundaries are statistically significant. The authors welcome suggestions for enhancements.

\section*{Supplementary Materials}
\label{sec:supplementary}
Data available from the Dryad Digital Repository: http://dx.doi.org/10.5061/dryad.2s201

\section*{Acknowledgments}
\label{Acknowledgments}

We thank Daril Vilhena, Shawn Laffan, and our colleagues and students for discussions. We also thank Anna Eklöf and two anonymous reviewers, associate editor James Albert, and chief editors Frank Andersson and Thomas Near for constructive comments on this manuscript. 

\section*{Funding}
\label{Funding}

This work was supported by the Swedish Research Council (B0569601 to A.A. and 2012-3729 to M.R.); the European Research Council under the European Union’s Seventh Framework Programme (FP/2007-2013 and ERC Grant Agreement n. 331024 to A.A.);  a Wallenberg Academy Fellowship to A.A.; and the S{\~a}o Paulo Research Foundation (2013/04170-8 and 2014/18837-7 to T.B.G.).


\end{document}